\documentclass[
aps, 
twocolumn
]{revtex4}

\usepackage[dvips]{graphicx}
\usepackage{amsmath}
\usepackage{amssymb}

\newcommand{\bra}[1]{\left\langle #1\right|}
\newcommand{\ket}[1]{\left| #1\right\rangle}

\newcommand{\ketbra}[2]{\left| #1\right\rangle\!\left\langle#2\right|}

\newcommand{\be}[0]{\begin{equation}}
\newcommand{\ee}[0]{\end{equation}}

\newcommand{\lra}\simeq
\newcommand{\eeqref}[1]{Eq.~(\ref{#1})}

\newcommand{\expec}[1]{\left\langle #1\right\rangle}

\begin{document}

\title{Undoing the effect of loss on quantum entanglement}

\author{Alexander E. Ulanov$^{1,2}$, Ilya A. Fedorov$^{1,3}$\footnote{The first two authors contributed equally to the work.}, Anastasia A. Pushkina$^{1,2,3}$, Yury V. Kurochkin$^1$, Timothy C. Ralph$^4$ and A. I. Lvovsky$^{1,3,5}$}

\affiliation{$^1$Russian Quantum Center, 100 Novaya St., Skolkovo, Moscow 143025, Russia}
\affiliation{$^2$Moscow Institute of Physics and Technology, 141700 Dolgoprudny, Russia}
\affiliation{$^3$P. N. Lebedev Physics Institute, Leninskiy prospect 53, Moscow 119991, Russia}
\affiliation{$^4$Centre for Quantum Computation and Communication Technology, School of Mathematics and Physics, University of Queensland, Brisbane, Queensland 4072, Australia}
\affiliation{$^5$Institute for Quantum Science and Technology, University of Calgary, Calgary AB T2N 1N4, Canada}

\email{LVOV@ucalgary.ca}
\date{\today}

\begin{abstract}
Entanglement distillation is a process via which the strength and purity of quantum entanglement can be increased probabilistically. It is a key step in many quantum communication and computation protocols. In particular, entanglement distillation is a necessary component of the quantum repeater, a device which counters the degradation of entanglement that inevitably occurs due to losses in a communication line. Here we report an experiment on distilling the Einstein-Podolsky-Rosen (EPR) state of light, the workhorse of continuous-variable entanglement, using the technique of noiseless amplification. In contrast to previous implementations, the entanglement enhancement factor achievable by our technique is not fundamentally limited and permits recovering an EPR state with a macroscopic level of entanglement no matter how   low the initial entanglement or how high the loss may be. In particular, we recover the original level of entanglement after one of the EPR modes has passed through a channel with a loss factor of 20. The level of entanglement in our distilled state is higher than that achievable by  direct transmission of any state through a similar loss channel. This is a key bench-marking step towards the realization of a practical continuous-variable quantum repeater and other CV quantum protocols.
\end{abstract}

%

\maketitle


\vspace{10 mm}

\section{Introduction.}
Quantum technology protocols exploit the unique properties of quantum systems to achieve communication, computing and metrology tasks that are impossible, inefficient or intractable for classical systems \cite{NIE00}. In many cases the distribution of entanglement, correlations between subsystems that exceed those possible for classical systems, is a necessary condition for quantum technology protocols to succeed. However, entanglement is fragile and can easily be degraded by the communication or storage of the entangled quantum systems. One solution to this problem is entanglement distillation \cite{BEN96}. Given an ensemble of weakly entangled quantum states, distillation techniques allow one to select or distill a smaller sub-ensemble of quantum states that are more strongly entangled. This can be achieved using only local operations and classical communication. In this way strong entanglement can be established between remote locations under conditions where it would be impossible without distillation --- such as losses that are common in quantum communication channels.

There are two broad classes of quantum optical technology protocols: those using quantum observables with a discrete spectrum, such as the spin of an electron, and those using quantum variables with a continuous spectrum, such as the position and momentum of a harmonic oscillator \cite{BRA05}. Our focus here is distillation of continuous-variable (CV) states. The primary entangled resource in CV systems is the two-mode squeezed vacuum state, also known as the Einstein-Podolsky-Rosen state \cite{REI89} because its idealized version was introduced by EPR in the early days of quantum mechanics to illustrate quantum nonlocality.

The EPR state can be used to implement many quantum protocols including continuous versions of teleportation and quantum key distribution \cite{WEE12}. An advantage of the CV approach to quantum communication is its universality: it is capable of transmitting arbitrary states of light, in contrast to the single-photon subspace of the Hilbert space to which the discrete method is limited. Further, homodyne detection used in CV systems does not require expensive single-photon detectors and is more tolerant to ambient light. Additionally, in contrast to their discrete-variable counterparts,  high-quality EPR states are readily available on demand at a high rate from parametric amplifiers.

On the other hand, those communication protocols that use single photons do not need a special distillation procedure to counter the effect of the losses. This is because if a photon is lost, it is not registered by the detector, so a loss event is automatically eliminated from further analysis. In CV protocols, quadrature detection occurs independently of the losses, so recovering an entangled resource suitable for use in a teleportation or repeater protocol requires a dedicated distillation step.

In this paper we present experimental results demonstrating the distillation of optical CV entanglement in two settings: for very low initial squeezing and after transmission through a lossy channel. In the second setting, we directly observe an entanglement strength of our distilled state that exceeds anything possible via deterministic transmission of the states through the same channel. That is, even if a perfectly pure, infinitely entangled EPR state was passed through that channel, the resulting entanglement would be inferior to what we observe for our distilled state. We will refer to this as breaking the deterministic bound.

Our protocol relies on the technique of noiseless linear amplification (NLA) \cite{RAL09}, in contrast to previous CV entanglement distillation demonstrations based on photon subtraction \cite{TAK10,KUR14}. Photon subtraction is unable to enhance entanglement in the EPR state by more than a factor of two, which is by far not sufficient to compensate for a loss occurring in a typical communication line. NLA does not suffer from this limitation, and in principle allows restoring the entanglement to a macroscopic level after an arbitrarily high loss \cite{BAR15}. It is this feature of NLA that enables us to break the deterministic bound. It represents a major step forward in realizing protocols that can enhance quantum technologies under practical conditions.

A key feature of our experiment is that heralded, free propagating distilled EPR states are produced by our protocol. This is in contrast to experiment \cite{CHR14}, where NLA has been emulated by postselecting the quadrature measurements of an asymmetrically attenuated EPR state. This virtual distillation protocol cannot herald free-propagating EPR pairs and hence cannot be used for quantum teleportation or repeaters.

\section{Theory.}
The principal scheme of our setup is shown in the inset to Fig.~\ref{f1}. We explain the idea of our method in the practically relevant case of low initial squeezing and high losses. The initial EPR state prepared in a pair of modes $A$ and $B$ is then given by
\begin{equation}\label{initstate}
\ket {\Psi_{AB}}=\ket{00}-\gamma\ket{11},
\end{equation}
where we have limited the analysis to the first order of the squeezing amplitude $\gamma\ll 1$. A loss channel with amplitude transmissivity $\tau\ll1$ applied to mode $B$ will degrade the non-vacuum component of this state, transforming it into a mixture
\begin{equation}
\hat\rho=(\ket{00}-\gamma \tau\ket{11})(\bra{00}-\gamma \tau\bra{11})+\gamma^2(1-\tau^2)\ketbra{10}{10}.
\end{equation}

We then subject this state to NLA. In contrast to previous NLA experiments \cite{XIA10,FER10,ZAV10,KOS13}, where either the method of generalized quantum scissors \cite{PEG98,BAB03} or coherent superposition of photon subtraction and addition was used, here we employ a technique known as quantum catalysis \cite{LVO02}. The mode to be amplified is reflected from a low-reflectivity beam splitter, on which it is brought into interference with an ancillary single photon. The amplification event is heralded by detecting a single photon in the other output port of that beam splitter (Fig.~\ref{f1}, inset). This will happen if either (a) the ancilla photon is reflected (probability amplitude of this event equals the amplitude reflectivity $r \ll 1$ of the beam splitter) or (b) a photon from the EPR state is transmitted (this occurs with probability amplitude $\gamma\tau$). In the latter case, the ancilla photon is likely to transmit through the beam splitter into the reflected channel of the EPR state. If these possibilities are made indistinguishable by proper mode matching of the ancilla photon and the EPR state, their amplitudes add coherently. As a result, the EPR state is transformed into
\begin{align*}
\hat\rho'&=(r\ket{00}-\gamma \tau\ket{11})(r\bra{00}-\gamma \tau\bra{11})\\&+r^2\gamma^2(1-\tau^2)\ketbra{10}{10}.
\end{align*}
The second term in the above state is of the higher order in $\gamma$, $\tau$, and $r$ than the second term, so the final state is well approximated by pure superposition
\begin{equation}\label{finalstate}
\ket {\Psi'_{AB}}=r\ket{00}-\gamma\tau\ket{11}.
\end{equation}
We see that the reduction of the single-photon component due to low initial squeezing and/or loss in the communication is compensated by the reduction of the vacuum part due to the catalysis. This is equivalent to NLA with an amplitude gain of $g=1/r$.

The amount of entanglement to be recovered depends on the ratio $r/\gamma\tau$ and reaches one ebit at $g=(\gamma\tau)^{-1}$. If we instead aim to optimize the two-mode squeezing of the resulting state, this goal is reached at a slightly lower gain of $g=[\gamma\tau(1+\sqrt 2)]^{-1}$ resulting in a two-mode squeezing factor of $2-\sqrt 2\approx 0.59$ with respect to the standard quantum limit. The precise value of the optimal gain may vary dependent on the efficiency of the ancilla photon source as well as the chosen continuous-variable entanglement measure. The bottom line is, though, that however high the loss, and however low the initial squeezing may be, one can choose the NLA gain to recover a macroscopic amount of entanglement in the distilled state \cite{SuppInfo}.

\section{Experiment.}
\begin{figure}[t]
	\includegraphics[width=\columnwidth]{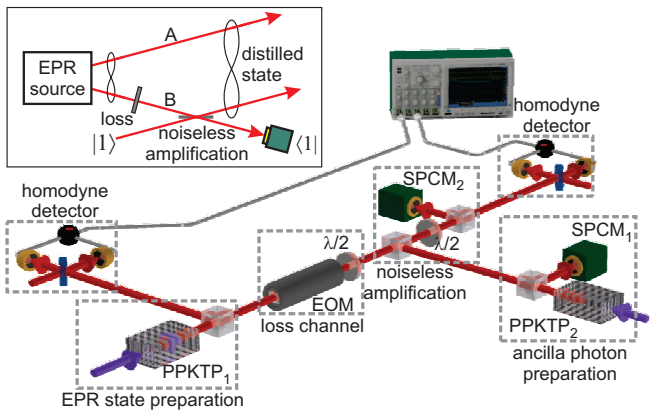}
	\caption{Experimental setup. Inset: principal scheme. The NLA distillation event corresponds to the click in the single-photon detector placed in the beam splitter output.}
	\label{f1}
\end{figure}

The EPR state (Fig.~\ref{f1}) is prepared in a periodically poled potassium titanyl phosphate  crystal (PPKTP$_1$), where the type II parametric down-conversion is taking place. The crystal is pumped with frequency-doubled pulses at $390$ nm, generated by a Ti:Sapphire laser with a repetition rate of 76 MHz and a pulse width of $\sim 1.8$ ps. We spatially split the outgoing orthogonally-polarized light modes on a polarizing beam splitter. Each mode is subjected to balanced homodyne detection \cite{Kumar2012}. The phase of one of the local oscillator phases is varied by a piezoelectric transducer (PZT). This allows us to obtain a quorum of the two-mode quadrature measurements and therefore completely characterize the state. We observe the variances in the difference and sum of the position quadratures of the initial state to be $\expec{(X_A-X_B)^2}=0.86\pm0.01$ and $\expec{(X_A+X_B)^2}=1.20\pm0.01$, measured in units of the vacuum state level [Fig.~\ref{f2}(a)]. This corresponds to the squeezing parameter of $\gamma=0.18\pm0.04$.

Prior to distillation, we degrade the initial state's entanglement by two alternative methods. First, we turn the polarization of the pump beam by $\theta=(76\pm 2)^\circ$, resulting in the reduction of the squeezing parameter $\gamma$ by a factor of $\cos\theta=0.22\pm0.02$.  This reduces the EPR correlations, bringing the difference and sum quadrature noise levels to $0.966$ and $1.044$, respectively. The second method consists of introducing asymmetric attenuation imitating propagation of mode 2 through a lossy optical channel with an intensity transmissivity of $\tau^2=0.05$. Then the difference and sum quadrature noise levels become $0.993$ and $1.010$, respectively. Such low squeezing levels could not be reliably measured by homodyne detection, so the variance values quoted above are based on the measured initial squeezing and the known attenuation factor.

\begin{figure*}
	\includegraphics[width=0.85\textwidth]{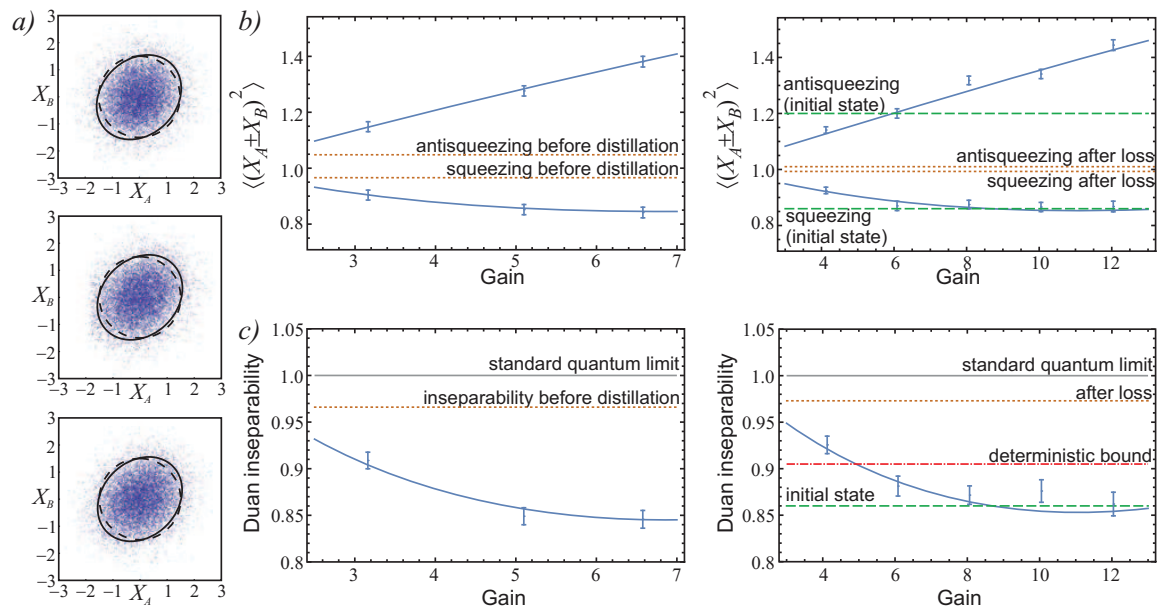}
	\caption{Experimental results. a) 10,000 samples of raw position quadrature data for the original two-mode squeezed (top) and the distilled state after applying the two methods of degrading the entanglement [reducing the initial squeezing (center), asymmetric loss (bottom)]. The NLA gain is $6.5$ for the center plot and 10 for the bottom. The degree of correlation is similar in all three cases. The quadratures exhibit nonclassical correlations at a level below the shot noise (dashed circle). b), c) Analysis of the distilled states as a function of the NLA gain. The left column in (b) and (c) corresponds to the case of reduced initial squeezing, the right column to the asymmetric loss. b) Two-mode squeezing measured by the variances of the sum (antisqueezing) and difference (squeezing) of the position quadratures in the two channels of the distilled EPR state. c) Inseparability parameter of Duan {\it et al.} \cite{DUA00}. The vertical axes in both (b) and (c) are scaled in units of standard quantum limit. The theoretical curves for the case of low initial squeezing are calculated assuming the initial squeezing parameter of $\gamma=0.05$, detection efficiency in the undistilled and distilled channels respectively $\eta_A=0.5$ and $\eta_B=0.5$, and the single-photon preparation efficiency $\eta=0.65$ \cite{KOS13,Berry1,Berry2}. For the case of a loss channel, $\gamma=0.135$ and $\eta_A=0.45$ are used; other parameters are the same \cite{SuppInfo}.}
	\label{f2}
\end{figure*}

The single photon state required for the NLA is obtained from parametric down-conversion in another crystal, PPKTP$_2$, at a rate of 50--100kHz, heralded by the clicks of a single-photon counting module (SPCM$_1$) \cite{HUI09}. The variable beam splitter that enacts the NLA is implemented by a combination of two polarizing beam splitters and a half-wave plate between them. One output of that beam splitter is directed onto another photon detector (SPCM$_2$) while the other one is subjected to homodyne detection. The distillation events correspond to both SPCMs firing synchronously, and occur at a rate of $\sim 10-100\mathrm{Hz}$.

To characterize the distilled state, the quadrature data from both detectors should be accompanied by knowledge of the sum $\varphi$ of the local oscillator phases for each data sample \cite{KUR14}. We acquire this information by utilizing the fraction of mode $B$ of the original EPR state that is reflected from the NLA beam splitter towards the homodyne detector in the absence of the heralding event. This field exhibits phase-dependent quadrature correlations with mode $A$. Because it is supplied at the master laser repetition rate, we observe a sufficiently clear correlation signal to retrieve $\varphi$ at any moment in time. It is the availability of this phase evaluation technique that has motivated us to choose the catalysis scheme in favor of the more common scissors scheme for the NLA.

To mitigate the degradation of this reference signal by the  loss channel, we implement that channel using an electro-optical modulator (EOM) followed by a half-wave plate. The EOM is controlled by a square wave of a 0.2 s period (during which the phase is assumed to vary linearly) and a duty cycle of 0.35. The ``off" period, when the transmission through the EOM is maximized, is used to measure $\phi$. During the ``on" period, the EOM voltage is set to attenuate the propagating signal, and the quadrature samples associated with the distilled EPR state are acquired.

\section{Results.} Our experimental results are summarized in Fig.~\ref{f2}. Part (a) shows samples of the raw data for the original and distilled states while the left and right columns in (b) and (c) correspond to the distilled states after applying the two methods of degrading the entanglement.

Panel (b) shows the variances of the sum and difference of the position quadratures (the behavior of the momentum quadrature is similar to that of the position). The amount of recovered squeezing is maximized at $g=$6--7 for the first method and $g=$10--12 for the second, which is consistent with the theoretical expectation in the ideal case $g=[\gamma\tau(1+\sqrt 2)]^{-1}$. In both cases, the two-mode squeezing of the state recovered by distillation is at the same level as the initial state, and dramatically higher than that of the EPR state with reduced squeezing, or the attenuated EPR state before distillation.

To quantitatively estimate the degree of entanglement at different stages of our experiment, we use the Duan {\it et al.} inseparability parameter \cite{DUA00}, defined as
\begin{equation}\label{}
I=\min\limits_a\frac{\expec{(X_A-aX'_B)^2}+\expec{(P_A+aP'_B)^2}}{1+a^2}.
\end{equation}
For all our data sets, we found the optimal value of $a$ to be close to unity, resulting in the plot shown in Fig.~\ref{f2}(c). The value of this parameter in the distilled state is a factor of about five greater than that in the degraded EPR states. To our knowledge, this factor of entanglement increase achieved in a distillation protocol is higher than anything reported to date, not only in the continuous-, but also in the discrete-variable domain \cite{KUR14,TAK10,PAN03}.

For distillation after asymmetric loss, we can also compare this parameter with that of the perfect EPR state that has experienced the same loss, in which case $I=(1-\tau^2)/(1+\tau^2)=0.905$ \cite{SuppInfo}. The entanglement observed in the distilled state is consistently stronger than this benchmark, demonstrating that our method is able to overcome the deterministic bound.

It is instructive to estimate our distillation procedure by finding a hypothetical ``equivalent" EPR state that, after appropriate attenuation of both channels, would exhibit the same quadrature statistics as our distilled state \cite{RAL11,SuppInfo}. We found the squeezing parameter of that state and the equivalent efficiency of the loss channel in mode $B$ to be, respectively, about two and three times greater than those in the actual experiment. This illustrates the capability of our procedure to largely compensate for the losses and low initial squeezing \cite{SuppInfo}. 

\section{Discussion and outlook.}
The CV entanglement distillation procedure demonstrated here allows us to compensate for the low initial squeezing and losses in the transmission channel, however significant these detrimental factors may be. For the first time, we have broken the deterministic bound for the distribution of CV entanglement. These properties make our method promising for the construction of a CV quantum error correction protocol \cite{RAL11} and hence ultimately a CV quantum repeater.

However, our results also demonstrate the limitations in the amount of entanglement that can be recovered. There are two reasons giving rise to these limitations. First, the single-channel linear-optical scheme implemented here constitutes a good approximation to NLA only for output states with mean photon numbers much less than one, thereby imposing a fundamental limit on the amount of  squeezing and entanglement attainable through the distillation. Second, we are not approaching even that limit because of imperfect preparation of the EPR and the ancilla photon, mode matching on the NLA beam splitter and detection inefficiencies.

The first of the above issues can be resolved by switching to multichannel NLA as envisioned in the original proposal \cite{RAL09}. Addressing the second point --- improving the experimental efficiencies --- is possible by a number of techniques. In particular, switching from the pulsed to continuous-wave domain would allow eliminating the intrinsic efficiency issues associated with pulsed squeezing \cite{PulsedSq1,PulsedSq2} and enable significant preparation and detection efficiency improvements \cite{HybridReview}. However, before these measures are to be undertaken, it is important to theoretically understand the requirements for a distillation device that would enable construction of a practical continuous-variable quantum repeater --- as was done for the discrete-variable case \cite{SCH11}. This understanding will clarify if and how the method presented here must be improved in order to qualify for practical applications.


\newpage
\makeatletter
\@addtoreset{equation}{section} 
\makeatother
\makeatletter
\@addtoreset{figure}{section} 
\makeatother
\makeatletter
\renewcommand{\thefigure}{S\@arabic\c@figure}
\makeatother
\section{Supplementary material}

\subsection{Duan's inseparability for the perfect EPR state}
Suppose a two-mode state prepared in modes $A$ and $B$ is subjected to one-sided loss in mode $B$. This loss results in quadrature transformation
\begin{align*}
X'_B&=\tau X_B+\sqrt{1-\tau^2}X_V;\\
P'_B&=\tau P_B+\sqrt{1-\tau^2}P_V,
\end{align*}
where $\tau$ is the amplitude transmissivity of the loss channel and $X_V$ is the position quadrature  of the vacuum state with variance $\expec{X_V^2}=1/2$.
We define the Duan {\it et al.} inseparability parameter \cite{DUA00} as
\begin{equation}\label{}
I=\min\limits_a\frac{\expec{(X_A-aX'_B)^2}+\expec{(P_A+aP'_B)^2}}{1+a^2}.
\end{equation}
The perfect EPR state has $X_A-X_B=0$ and $P_A+P_B=0$, but each quadrature taken individually has infinite variance. Because of the latter fact, the only value of $a$ for which $I$ takes a finite value is $a=1/\tau$. Then for the position quadratures we have
\begin{align*}\expec{(X_A-aX'_B)^2}&=\expec{\left(X_A-X_B-\frac{\sqrt{1-\tau^2}}{\tau}X_V\right)^2}\\&=\frac 12\frac{1-\tau^2}{\tau^2}.\end{align*}
The same value obtains for the momentum term: $\expec{(P_A+aP'_B)^2}=\frac 12\frac{1-\tau^2}{\tau^2}.$ Accordingly, we have
$I=\frac{1-\tau^2}{1+\tau^2}.$ For our experiment, $\tau^2=0.95$ and hence $I=0.905$.

\subsection{Theoretical description}
\paragraph{Ideal case}
We begin by considering the idealized case described by Eq.~(3) in the main text. The variances of the individual position or momentum quadratures in that state are
\begin{eqnarray}\label{covar1}
\expec{X_A^2}&=&\expec{X_B^2}=\expec{P_A^2}=\expec{P_B^2}\\
&=&\frac{r^2+3(\gamma \tau)^2}{2[r^2+(\gamma \tau)^2]}\nonumber
\end{eqnarray}
while their correlation equals
\begin{eqnarray}
-\expec{X_AX_B}&=&\expec{P_AP_B}\label{covar2}\\
&=&\frac{r\gamma \tau}{r^2+(\gamma \tau)^2}.\nonumber
\end{eqnarray}
The variance of the quadrature sum and difference is therefore
$$\expec{(X_A\pm X_B)^2}=\expec{(P_A\mp P_B)^2}=\frac{\beta^2+3\pm2\beta}{\beta^2+1},$$
where $\beta=\frac{r}{\gamma \tau}=\frac 1{g\gamma \tau}$, $g=\frac 1 r$ being the NLA gain.
The position difference (momentum sum) variance is minimized for $\beta=1+\sqrt 2$, at which point it takes the value $\expec{(X_A- X_B)^2}=\expec{(P_A+ P_B)^2}=2-\sqrt 2$.

\begin{figure}[t]
	\includegraphics[width=0.8\columnwidth]{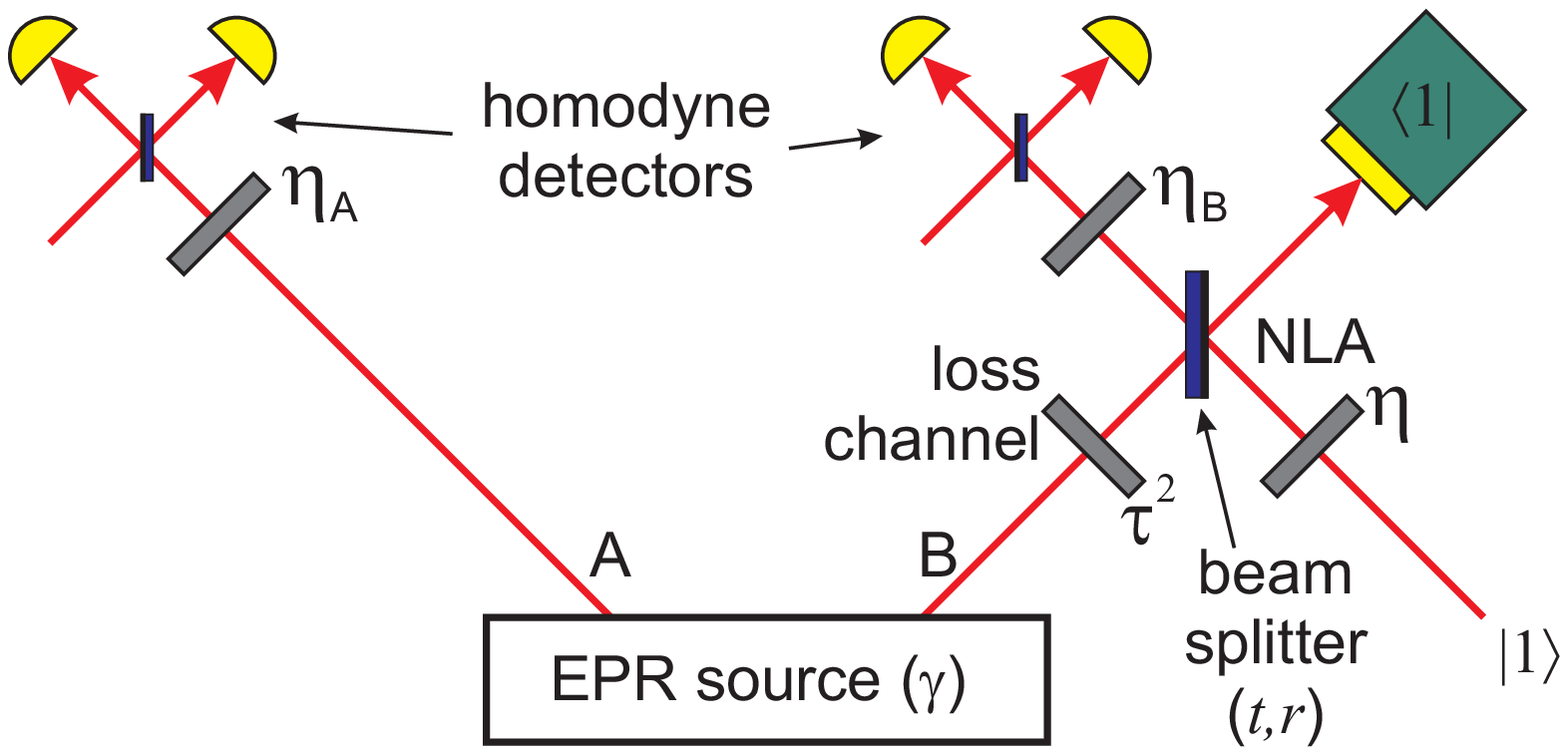}
	\caption{The model for a complete theoretical analysis. Grey rectangles indicate attenuators, quantities next to them show intensity transmissivities of these attenuators.}
	\label{s1}
\end{figure}

\paragraph{Treatment at the single-photon level}
Let us now develop a more complete theory, taking into account the efficiency $\eta$ of the ancilla photon as well as the efficiencies $\eta_A$ and $\eta_B$ of the detectors in both channels, according to Fig.~S1. Our goal is to calculate the covariance matrix of the distilled state. We continue to work in the subspace of the Hilbert space spanned by states $\ket 0, \ket 1$ and assume the perfect efficiency of the heralding SPCM. Both these assumptions are well-justified for $\gamma,\tau\ll 1$ and $r\sim\gamma \tau$. The inefficient ancilla photon source generates the mixture of the single-photon state with probability $\eta$ and the vacuum with probability $1-\eta$. In the former case, the distilled EPR state is $r\ket{00}-\gamma\tau\ket{11}$, while in the latter case, the click of the heralding detector will occur only if the photon from the EPR state propagates through the beam splitter towards the SPCM, resulting in the unnormalized state $\gamma \tau\ket{10}$ in the distilled output. Taking into account both possibilities, we obtain the following output state
\begin{eqnarray}\label{rhooutprime}\hat\rho'&=&\eta(r\ket{00}-\gamma\tau\ket{11})(r\bra{00}-\gamma\tau\bra{11})\\
&+&(1-\eta)(\gamma \tau)^2\ketbra{10}{10},\nonumber
\end{eqnarray}
where we have not yet taken into account the homodyne detector efficiencies $\eta_A$ and $\eta_B$.

\begin{figure}[t]
	\includegraphics[width=0.8\columnwidth]{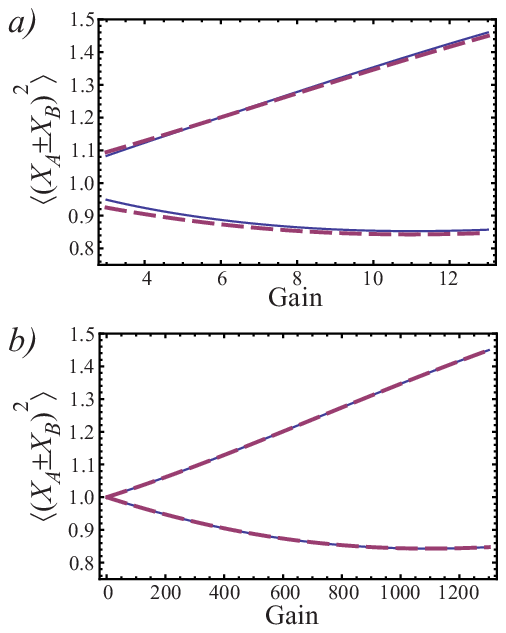}
	\caption{Comparison between the prediction of the full theory (thin solid blue curves) and that at the single-photon level (thick dashed purple curves). a) Same parameters as used in the experimental fit: $\tau^2=0.05,\gamma=0.135,\eta_A=0.45,\eta_B=0.5,\eta=0.65$. b) Same parameters as in (a) except $\tau^2=0.0005,\gamma=0.0135$. In (b), a much lower multiphoton fraction results in a better match between the two theories.}
	\label{f2}
\end{figure}

In the second term of the above equation we have
\begin{align*}
\expec{X_A^2}&=\expec{P_A^2}=\frac 3 2;\\
\expec{X_B^2}&=\expec{P_B^2}=\frac 1 2;\\
\expec{X_AX_B}&=\expec{P_AP_B}=0.
\end{align*}
From these results, as well as from Eqs.~\eqref{covar1} and \eqref{covar2}, we find the covariance matrix elements of state \eqref{rhooutprime}:
\begin{align*}
\expec{X_A^2}_{\hat\rho'}&=\expec{P_A^2}_{\hat\rho'}=\frac{\eta(\beta^2+3)+3(1-\eta)}{2(1+\eta\beta^2)};\\
\expec{X_B^2}_{\hat\rho'}&=\expec{P_B^2}_{\hat\rho'}=\frac{\eta(\beta^2+3)+(1-\eta)}{2(1+\eta\beta^2)};\\
-\expec{X_AX_B}_{\hat\rho'}&=\expec{P_AP_B}_{\hat\rho'}=\frac{\eta\beta}{1+\eta\beta^2}.
\end{align*}
Accounting for the detector efficiencies affects these results as follows:
\begin{eqnarray}
\expec{X_A^2}_{out}&=&\eta_A\expec{X_A^2}_{\hat\rho'}+\frac 1 2(1-\eta_A); \nonumber\\
\expec{X_B^2}_{out}&=&\eta_B\expec{X_B^2}_{\hat\rho'}+\frac 1 2(1-\eta_B); \label{losscovar}\\
\expec{X_AX_B}_{out}&=&\sqrt{\eta_A\eta_B}\expec{X_AX_B}_{\hat\rho'}.\nonumber
\end{eqnarray}
The covariance matrix elements related to the momentum are transformed in a similar fashion.

\paragraph{Full theory}
In addition to the above analytical calculation, we performed a numerical modeling of our experiment using the density matrix in the Fock basis, in the subspace spanned by the zero- to three-photon Fock states. Figure S2 shows the variances $\expec{(X_A\pm X_B)^2}$ predicted by the two techniques. As expected, the predictions are very similar as long as conditions $\gamma,\tau\ll 1$ and $r\sim\gamma \tau$ hold. The single-photon approximation breaks down for low amplifications because for macroscopic NLA beam splitter reflectivities $r\lesssim 1$ we obtain a significant likelihood of Hong-Ou-Mandel type events when both the ancilla photon and the photon from the EPR state exit the beam splitter through the same port.

We used this numerical model to calculate the theoretical fit to the data shown in Fig.~2 of the main text. Note that our model does not take into account a possible mode mismatch between the to-be amplified  EPR state mode and the ancilla photon. This is a possible explanation for the discrepancy between the measured and best-fit values for the initial squeezing ($\gamma=0.18\pm 0.04$ and $\gamma=0.135$, respectively).
 \begin{figure}[t]
	\includegraphics[width=0.8\columnwidth]{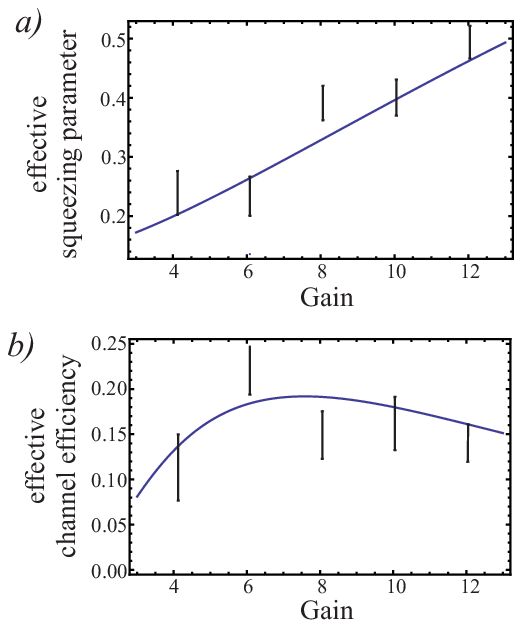}
	\caption{The effective equivalent parameters of the distilled EPR state: channel efficiency (a) and squeezing parameter (b), as explained in the text.}
	\label{f3}
\end{figure}

\subsection{Calculating the equivalent EPR state}
The variances of the sum and difference of the position quadratures of a pure EPR state with squeezing parameter $\gamma_{eq}$ are $e^{\pm2\gamma_{eq}}$, respectively \cite{SqReview}. If this state experiences two-sided loss by propagating through channels with efficiencies $\eta_{A,eq}$ and $\eta_{B,eq}$, covariance matrix of the output can be calculated from \eeqref{losscovar}:
\begin{align*}
\expec{X_A^2}_{eq}&=\expec{P_A^2}_{eq}=\frac 1 2[\eta_{A,eq}\cosh 2\gamma_{eq}+(1-\eta_{A,eq})];\\
\expec{X_B^2}_{eq}&=\expec{P_B^2}_{eq}=\frac 1 2[\eta_{B,eq}\cosh 2\gamma_{eq}+(1-\eta_{B,eq})];\\
-\expec{X_AX_B}_{eq}&=\expec{P_AP_B}_{eq}=\frac 12\sqrt{\eta_{A,eq}\eta_{B,eq}}\sinh 2\gamma_{eq}.
\end{align*}
Accordingly, the variances of the sum and difference of the position quadratures become
\begin{eqnarray}
\expec{(X_A\pm X_B)^2}_{eq}&=&1+\frac{\eta_{A,eq}+\eta_{B,eq}}2(\cosh 2\gamma_{eq}-1)\nonumber \\
&\mp&\sqrt{\eta_{A,eq}\eta_{B,eq}}\sinh 2\gamma_{eq}.\label{equivvar}
\end{eqnarray}
Knowing the experimentally measured quantities $\expec{(X_A\pm X_B)^2}$ for the distilled state, one can solve the above equation to find the parameters of the equivalent state that would exhibit similar quadrature statistics. In the calculations shown in Fig.~\ref{f3}, we assumed the efficiency $\eta_{A,eq}$ of the equivalent state to equal the actual efficiency $\eta_A$ of the undistilled channel obtained from the fit to the experimental data, and calculated $\gamma_{eq}$ and $\eta_{B,eq}$. In this way, the quantity  $\eta_{B,eq}$ describes the equivalent channel efficiency, thereby serving as a benchmark of our distillation protocol.

The high uncertainty bars in Fig.~\ref{f3} are due to the relatively low degree of squeezing obtained after the distillation. Small variations of quadrature variance measurements have significant effect on the solutions of \eeqref{equivvar}.


\begin{thebibliography}{99}

\bibitem{NIE00} M. Nielsen and I. Chuang, {\it Quantum computation and quantum information}, (Cambridge University Press, Cambridge, UK 2000).

\bibitem{BEN96} C. Bennett {\it et al.}, Purification of noisy entanglement and faithful teleportation
via noisy channels. Phys. Rev. Lett. {\bf 76}, 722
(1996).

\bibitem{BRA05}
S. L. Braunstein and P. van Loock, Quantum information with continuous variables,
Rev. Mod. Phys. {\bf 77}, 513 (2005).

\bibitem{REI89} M. D. Reid, Demonstration of the Einstein-Podolsky-Rosen paradox using
nondegenerate parametric amplification. Phys. Rev. A {\bf 40}, 913
(1989).

\bibitem{WEE12} C. Weedbrook, S. Pirandola, R. Garcia-Patron, N. J. Cerf, T. C. Ralph, J. H. Shapiro, S. Lloyd, Gaussian Quantum Information,
Rev. Mod. Phys. {\bf 84}, 621 (2012).



\bibitem{RAL09}
T. C. Ralph, and A. P. Lund,  Nondeterministic Noiseless Linear Amplifcation of Quantum Systems, Quantum Communication Measurement and Computing, Proceedings
of 9th International Conference, Lvovsky, A. (editor), 155--160 (AIP, New York 2009).

\bibitem{TAK10}
H. Takahashi, J. S. Neergaard-Nielsen, M. Takeuchi, M. Takeoka, K. Hayasaka, A. Furusawa, M. Sasaki, Non-Gaussian entanglement distillation for continuous variables,
Nature Photonics {\bf 4}, 178 (2010).

\bibitem{KUR14} Y. Kurochkin, A. S. Prasad,  and A. I. Lvovsky, Distillation of the two-mode squeezed state, Phys. Rev. Lett. {\bf 112}, 070402 (2014).

\bibitem{BAR15} T. J. Bartley and I. A. Walmsley, Directly comparing entanglement-enhancing non-Gaussian operations, New J. Phys. {\bf 17}, 023038 (2015).

\bibitem{CHR14}
H. M. Chrzanowski, N. Walk, S. M. Assad, J. Janousek, S. Hosseini, T. C. Ralph, T. Symul, and P. K. Lam, Measurement-Based Noiseless Linear Amplification for Quantum Communication, Nature Photonics {\bf 8}, 333 (2014).

\bibitem{XIA10}  G. Y. Xiang, T. C. Ralph, A. P. Lund, N. Walk and G. J. Pryde, Noiseless Linear Amplification and Distillation of Entanglement, Nature Photonics {\bf 4}, 316 (2010).

\bibitem{FER10}  F. Ferreyrol, et al., Implementation of a nondeterministic optical noiseless
amplifier, Phys. Rev. Lett. {\bf 104}, 123603 (2010).

\bibitem{ZAV10} A. Zavatta, J. Fiurasek, and M. Bellini, A high-fidelity noiseless amplifier for
quantum light states. Nature Photon. {\bf 5}, 52
(2010).

\bibitem{KOS13} S.Kosis, G.Y.Xiang, T.C.Ralph and G.J.Pryde, Heralded Noiseless Amplification of a Photon Polarization Qubit, Nature Physics {\bf 9}, 23 (2013)

\bibitem{PEG98} D. Pegg, L. Phillips, S. Barnett, Optical state truncation by projection
synthesis. Phys. Rev. Lett. {\bf 81}, 1604
(1998).

\bibitem{BAB03} S. A. Babichev, J. Ries,  and A. I. Lvovsky, Quantum scissors: teleportation of single-mode optical states by means of a nonlocal single photon, Europhys. Lett. {\bf 64}, 1 (2003).

\bibitem{LVO02} A. I. Lvovsky, J. Mlynek, Quantum-Optical Catalysis: Generating Nonclassical States of Light by Means of Linear Optics, Phys.Rev.Lett. {\bf 88}, 250401 (2002).


\bibitem{SuppInfo} See the Supplementary information for the details of all theoretical calculations.

\bibitem{Kumar2012}
R. Kumar, E. Barrios, A. MacRae, E. Cairns, E.H. Huntington, and A.I. Lvovsky, Opt. Commun., {\bf 285}, 5259 (2012).

\bibitem{HUI09} S. R. Huisman, N. Jain, S. A. Babichev, F. Vewinger, A.-N. Zhang, S.-H. Youn, and A. I. Lvovsky, Instant single-photon Fock state tomography, Opt. Lett. {\bf 34}, 2739 (2009).

\bibitem{DUA00} L.-M. Duan, G. Giedke, J. I. Cirac, and P. Zoller, Inseparability Criterion for Continuous Variable Systems, Phys. Rev. Lett. {\bf 84}, 2722 (2000).

\bibitem{PAN03} J. W. Pan, S. Gasparoni, R. Ursin, G. Weihs, A. Zeilinger, Experimental entanglement purification of arbitrary unknown states, Nature {\bf 423}, 417 (2003).

\bibitem{RAL11} T. C. Ralph, Quantum error correction of continuous-variable states against Gaussian noise, Phys. Rev. A {\bf 84}, 022339 (2011).

\bibitem{PulsedSq1} W. Wasilewski, A. I. Lvovsky, K. Banaszek, and C. Radzewicz, Pulsed squeezed light: simultaneous squeezing of multiple modes
Phys. Rev. A {\bf 73}, 063819 (2006).

\bibitem{PulsedSq2} A. I. Lvovsky, W. Wasilewski and K. Banaszek, Decomposing a pulsed optical parametric amplifier into independent squeezers,
J. Mod. Opt. {\bf 54}, 721 (2007).

\bibitem{HybridReview} U. L. Andersen, J. S. Neergaard-Nielsen, P. van Loock, A. Furusawa,   Hybrid quantum information processing, arXiv:1409.3719 (2014).

\bibitem{SCH11} A. Scherer, B. C. Sanders, W. Tittel, Opt. Express {\bf 19}, 3004 (2011).

\bibitem{Berry1} D. W. Berry and A. I. Lvovsky, Linear-optical processing cannot increase photon efficiency, Phys. Rev. Lett. {\bf 105}, 203601 (2010).

\bibitem{Berry2} D. W. Berry and A. I. Lvovsky, Preservation of loss in linear-optical processing,
Phys. Rev. A {\bf 84}, 042304 (2011).







\end{thebibliography}

\begin{thebibliography}{99}
\bibitem{DUA00} L.-M. Duan, G. Giedke, J. I. Cirac, and P. Zoller, Inseparability Criterion for Continuous Variable Systems, Phys. Rev. Lett. {\bf 84}, 2722 (2000).
\bibitem{SqReview} A. I. Lvovsky, Squeezed light, arXiv:1401.4118.
\end{thebibliography}
\end{document}